# Real-time ratiometric optical nanoscale thermometry

Yongliang Chen,[1] Chi Li,[1] Tieshan Yang,[1] Evgeny A. Ekimov,[2,3] Carlo Bradac,[4] Milos Toth,[1,5] Igor Aharonovich[1,5] and Toan Trong Tran[1,*]

1. School of Mathematical and Physical Sciences, University of Technology Sydney, Ultimo, NSW, 2007, Australia
2. Institute for High Pressure Physics, Russian Academy of Sciences, Troitsk 108840, Moscow, Russia
3. Lebedev Physics Institute, Russian Academy of Sciences, Moscow 117924, Russia
4. Trent University, Department of Physics & Astronomy, 1600 West Bank Drive, Peterborough, ON, K9L 0G2
5. ARC Center of Excellence for Transformative Meta-Optical Systems (TMOS), Faculty of Science, University of Technology Sydney, Ultimo, NSW, 2007, Australia

*Corresponding author: trongtoan.tran@uts.edu.au

**ABSTRACT**

All-optical nanothermometry has become a powerful, non-invasive tool for measuring nanoscale temperatures in applications ranging from medicine to nano-optics and solid-state nanodevices. The key features of any candidate nanothermometer are sensitivity and (signal, spatial and temporal) resolution. Here, we demonstrate a real-time, diamond-based nanothermometry technique with sensitivity (0.018 K$^{-1}$) and resolution (5.8×10$^4$ K·Hz$^{-1/2}$·W·cm$^{-2}$) much larger than those of any existing all-optical method. The distinct performance of our approach stems from two factors. i) Temperature sensors—nanodiamonds co-hosting two Group IV colour centers—engineered to emit spectrally-separated Stokes and Anti-Stokes fluorescence signals under excitation by a single laser source. ii) A parallel detection scheme based on filtering optics and high-sensitivity photon counters for fast readout. We demonstrate the performance of our method by monitoring temporal changes in the local temperature of a microcircuit and a MoTe$_2$ field-effect transistor. Our work lays the foundation for time-resolved temperature monitoring and mapping of micro-/nano-scale devices such as microfluidic channels, nanophotonic circuits, and nanoelectronic devices, as well as complex biological environments such as tissues and cells.

Research in nanoscale optical thermometry has recently gained tremendous momentum owing to its ability to determine, non-invasively, the temperature of microscopic objects with high sensitivity and high spatial resolution.[1-5] This has made nanothermometry a powerful tool for applications in several areas of research, including medicine,[6] biophotonics,[4] and solid-state nanoelectronics.[7-10] The latter is a prime example of how important nanothermometry has become, for the increasing miniaturization of integrated electronics renders heat dissipation a major problem in micro- and nano-devices. At these scales, heat transport is heavily restricted and ceases to obey Fourier's law[11]—which is instead widely applicable to meso- and macro-scopic systems. Highly-localised, thermal hot spots arise frequently in nanoscale electronic devices and can result in detrimental phenomena such as overheating, loss of performance and catastrophic failure.[12] The ability to monitor temperature at the nanoscale is thus of critical importance for developing high-performance computing devices and energy-efficient electronics. Moreover, nanothermometry has found widespread application in life sciences, as temperature plays a key role in cellular metabolic processes,[13,14] and is showing potential for the design of target-specific diagnostic and therapeutic strategies, such as hyperthermia and immunotherapy.[6,15-17]

The key constituent of any nanothermometry approach is the sensor. In optical methods, this is usually a fluorescent nanoprobe that reacts to changes in temperature by variations in its fluorescence characteristics. The variety of available optical nanothermometers is large and includes fluorescent dyes,[1,2] upconversion nanoparticles,[4,18] semiconductor quantum dots,[19-21] plasmonic nanoparticles,[22] nanodiamonds[23,24] and hBN nanoflakes.[25] Among these systems, fluorescent nanodiamonds have emerged as some of the most promising thermosensors owing to a combination of appealing properties. They host bright, photostable color centres that are sensitive to temperature; they are biocompatible, while also able to withstand extreme temperatures and harsh chemical environments.[26-28] In typical diamond-based nanothermometry, temperature changes are detected by measuring corresponding variations in the fluorescence spectral features of the color centers. These include measuring either shifts in the resonance frequencies of optically detected magnetic resonance (ODMR) spectra,[29] or measuring spectral changes in the zero-phonon line (ZPL) emission wavelength,[30] ZPL linewidth[31] or emission intensity.[32] Approaches based on ODMR frequency shifts of the negatively-charged nitrogen-vacancy (NV⁻) center have achieved some of the highest resolutions (~9 mK Hz$^{-1/2}$) in diamond nanothermometry. However, the method suffers from a series of drawbacks. It requires excitation of the nanosensors with microwave (MW) radiation which may induce heating. Moreover, even extremely small, non–quasi-static magnetic and electric fields can cause shifts in the NV spin resonance, which are indistinguishable from those induced by temperature changes and are thus a source of undesired thermal equivalent noise (TEN), unless decoupling spin-echo techniques are used. Finally—and most importantly in this context—these measurements are relatively slow and require relatively high laser excitation powers.

Partially due to these drawbacks, all-optical nanothermometry techniques based on the fluorescence of diamond color centers have gained attention. For these, both excitation of the centers and detection of their fluorescence are optical, making them highly versatile and attractive for a variety of real-world realizations, especially in biology and nanoelectronics. However, most of these approaches rely on the monitoring and subsequent analysis of target spectral features (ZPL, linewidth, intensity, etc.), which can lead to limited throughput and/or temperature resolution.

A highly-appealing nanothermometry approach employs the intensity ratio of the Stokes emission and the phonon-mediated Anti-Stokes emission of a group IV color center in diamond.[32,33] When first demonstrated, this method showed sensitivities higher than those of other all-optical methods, with comparable resolutions. However, it requires the consecutive excitation of the colour center with two separate lasers, at wavelengths above and below the emission wavelength. In addition, as most all-optical approaches, it relies on spectral analysis, which causes the aforementioned limits in throughput and/or resolution. Here, we demonstrate a nanothermometry technique that solves these problems and achieves a 2-fold improvement in sensitivity and a 30-fold improvement in resolution relative to power density, over the next best all-optical methods. The method is based on measuring the Anti-Stokes/Stokes intensity ratio, albeit with two key differences. First, we remove the need for consecutive excitation with two lasers by using nanodiamonds co-doped with Si and Ge. The nanodiamonds host both silicon-vacancy (SiV) and germanium-vacancy (GeV) centers, which are spectrally separated and are simultaneously excited by a single laser in the Stokes and Anti-Stokes regime, respectively. Second, we perform the readout of the SiV and GeV emission intensities in parallel using two photodetectors rather than a spectrometer. This enables the real-time readout of the fluorescence intensity ratio and completely eliminates the need for spectral analysis. Moreover, these benefits are realised whilst achieving a temperature sensitivity and resolution that outperform all existing all-optical nanothermometry techniques. This is important since these metrics ultimately limit the usefulness and applicability of any high-speed nanothermometry technique. Our method is highly attractive for fast monitoring of temperature variations at the nanoscale, as we demonstrate here using a microcircuit and a multilayer $MoTe_2$ field-effect transistors (FET).

The two types of color centers were incorporated in nanodiamonds using the high-pressure high-temperature (HPHT) growth technique (c.f. Methods).[34] Briefly, a powder mixture of adamantane ($C_{10}H_{16}$), tetraphenylgermane ($C_{24}H_{20}Ge$) and tetraphenylsilane ($C_{24}H_{20}Si$) were pressed into a pellet and placed inside a titanium capsule. The capsule was then compressed at high temperature (1800–2000 °C) and high pressure (8–9 GPa) inside a reaction chamber, and cooled under high pressure to room temperature. The resulting diamond particles were then dispersed in isopropanol (IPA), drop-casted on a clean silicon substrate and left to dry on a hotplate at 60 °C to completely remove the residual solvent. Two representative scanning electron microscopy (SEM) images showing diamonds of different sizes are displayed in **Supplementary Figure S1a**. The sizes of these diamond particles ranged from a few micrometres to a few hundred nanometers and can be controlled by adjusting the parameters in the synthesis process. By using Raman spectroscopy, we observed the characteristic narrow diamond peak at ~1332 cm$^{-1}$, (**Supplementary Figure S1b**) indicating that the particles were high-quality diamonds.

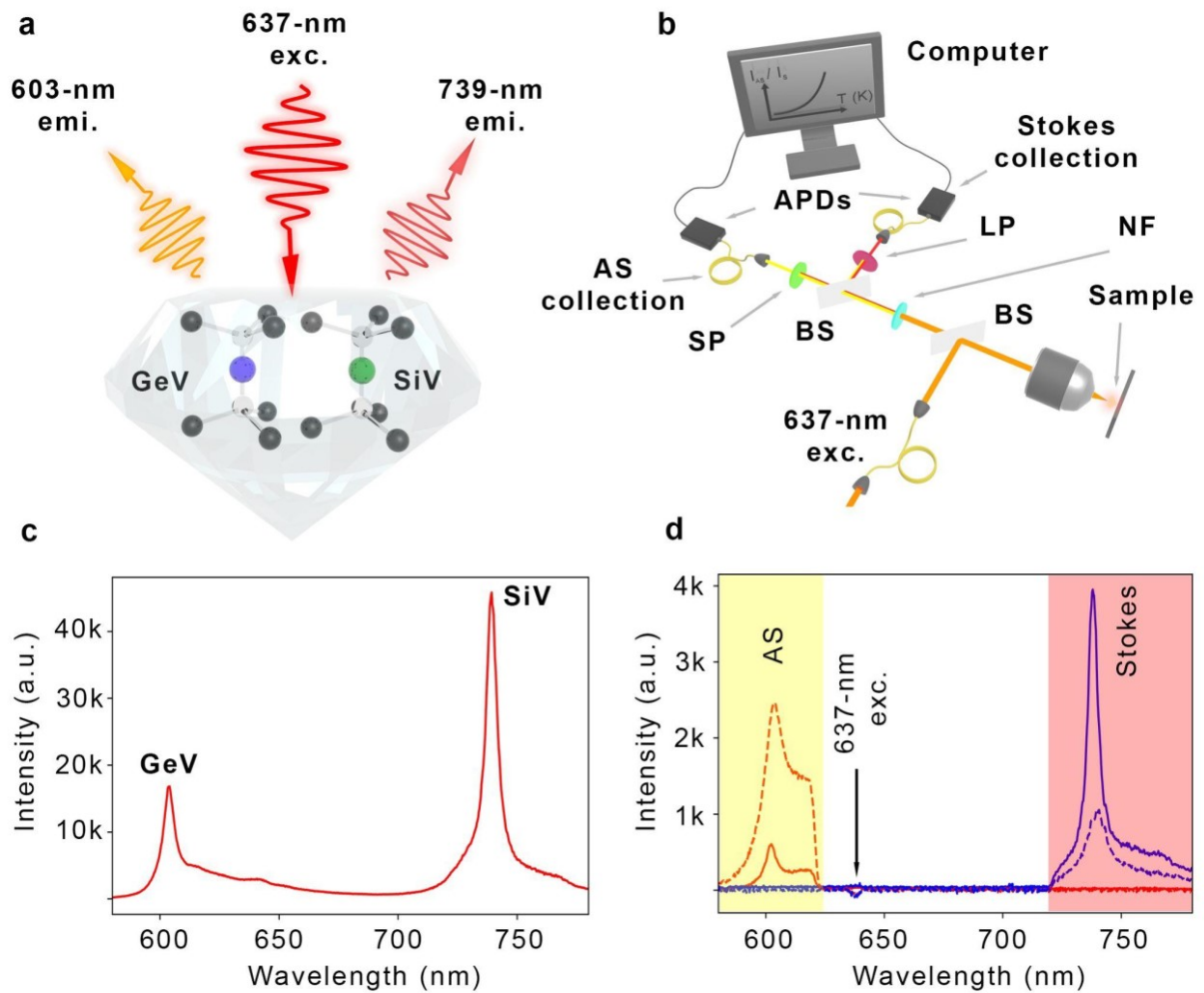

**Fig. 1| Anti-Stokes and Stokes photoemission from GeV and SiV centres in a nanodiamond generated by a single excitation source. a**, Schematic showing a 637-nm laser (red) used to excite the GeV-SiV co-doped diamonds: Anti-Stokes regime for GeV centers (emitting at 603 nm) and Stokes regime for SiV centers (emitting at 739 nm), respectively. **b**, Schematic of the optical setup. The fluorescence signals from the GeV and SiV centers are split, filtered and collected by two avalanche photodiodes (APDs) on separate arms. BS: beam-splitter; NF: neutral-density filter; LP: long-pass filter; SP: short-pass filter; AS: Anti-Stokes. **c**, Characteristic spectrum showing both the GeV and SiV emission peaks. Here, both centers were excited in the Stokes configuration, using a 532-nm, 300-$\mu$W laser. **d**, Two representative spectra obtained with a single 637-nm excitation source at room-temperature (solid lines) and 130 °C (dashed lines), displaying both Anti-Stokes (GeV) and Stokes fluorescence (SiV). The transparent yellow and red shadings indicate the filtered bands used for temperature sensing. The acquisition times in **c** and **d** are 1 s and 150 s, respectively.

**Figure 1a** shows the proposed scheme where a 637-nm laser (red arrow) excites the GeV center (Anti-Stokes) and SiV center (Stokes), producing emission at ~603-nm (orange arrow) and ~739-nm (dark red arrow), respectively. The 637-nm wavelength for the excitation laser was chosen because it lies within the phonon sideband (PSB) of GeV centers and had been previously demonstrated to efficiently excite these centers in the Anti-Stokes regime. Any wavelength in the range ~625–670 nm can be used, as in this range both GeV and SiV centers are excited.

To characterize the behavior of the co-doped diamonds, we use a lab-built confocal microscope shown in **Figure 1b** (c.f. Methods). The fluorescence signal from the diamond nanoparticles is split into two collection arms where two separate fiber-coupled avalanche photodiodes (APDs) collect the filtered signal from either the GeV or SiV centers. This allows us to obtain the real-time GeV (Anti-Stokes)/SiV(Stokes) fluorescence intensity ratio which is directly translated into a temperature value using a calibration curve (see below). This eliminates completely the need to collect spectra and to perform post-processing steps conventionally used in analogous methods.

**Figure 1c** displays the main spectral properties of the engineered nanodiamonds co-doped with GeV and SiV centers. The spectrum in the figure was obtained using a continuous-wave 532-nm excitation laser, at room-temperature. Two prominent sharp peaks are clearly visible; they are the characteristic ZPL wavelengths of the GeV (emission at ~603 nm) and the SiV centers (emission at ~739 nm). To verify the feasibility of the proposed nanothermometry technique, we then excited the same nanodiamond with a 637-nm laser, which should excite the GeV and the SiV centers in the Anti-Stokes and Stokes regime, respectively. We conducted the measurement at two different temperatures as shown in **Figure 1d**. At room-temperature, the intensity of the GeV peak (~603 nm) was almost 8 times smaller than that of the SiV peak (~739 nm). At 130 °C, however, the intensity of the GeV peak was nearly 2.5 times higher than that of the SiV peak. Such a drastic change in the intensity ratio of GeV to SiV peak with temperature indicates that the ratio is suitable for high-sensitivity thermometric measurements. The changes in fluorescence of the GeV and SiV centers are attributed to competing mechanisms. Briefly, an increase in temperature causes a decrease in fluorescence due primarily to the activation of non-radiative decays from the excited state.[35,36] This behaviour is the main factor determining the temperature response of SiV centers, which are excited in the Stokes regime. While this mechanism is still present for the GeV centers, their fluorescence response to temperature changes is dominated by the temperature dependence of the phonon spectral density, i.e., the phonon density of states multiplied by the transition amplitude. This is because the GeV centers are excited in the Anti-Stokes regime, with a low-energy (long-wavelength) laser. Their optical transition to the first excited electronic state thus depends on them being in an excited vibronic state—populated via the absorption of phonons by ground-state electrons—and depends strongly (in fact, exponentially) on temperature.[32]

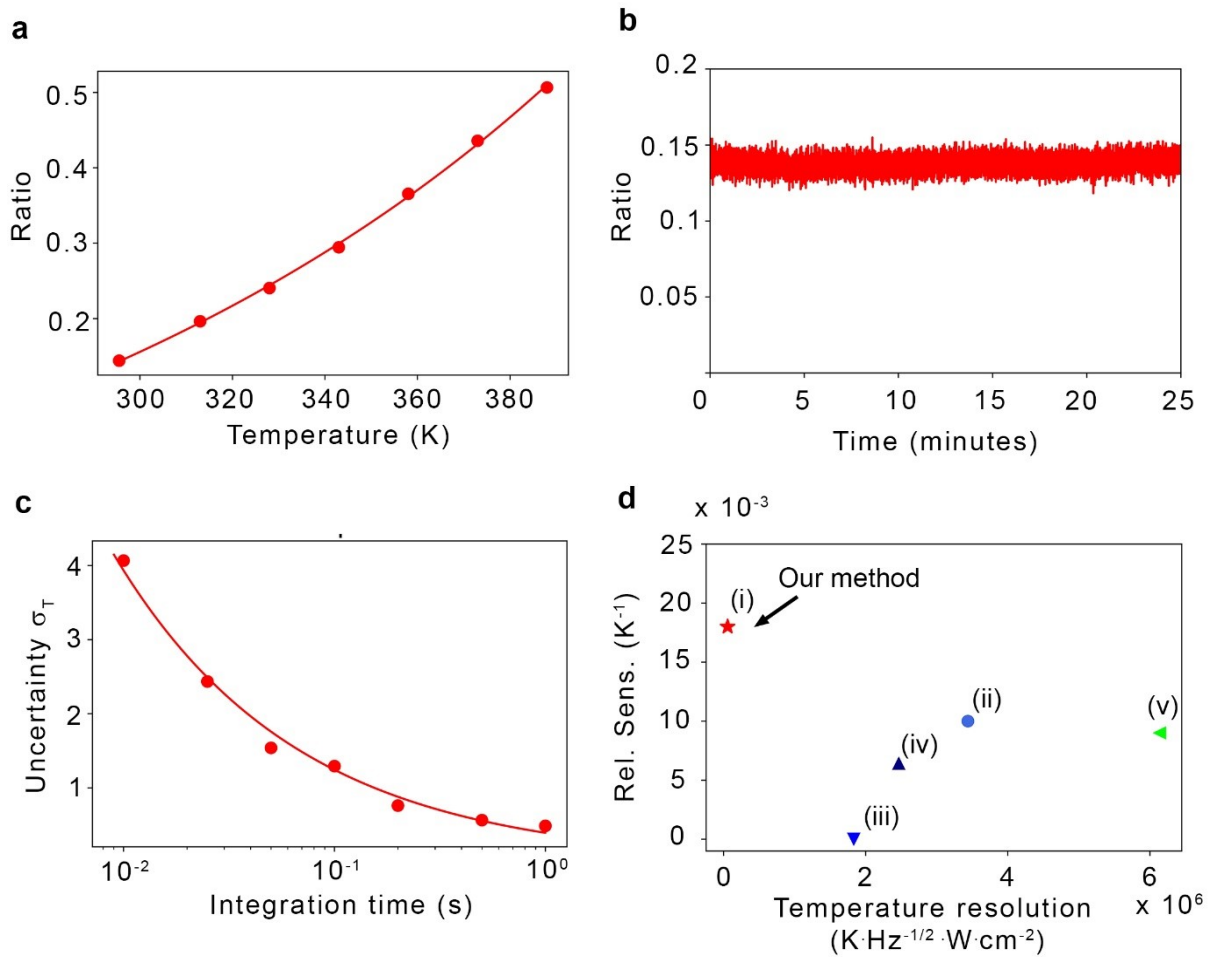

**Fig. 2| Nanothermometry based on the GeV/SiV Anti-Stokes/Stokes emission ratio. a**, Temperature dependence of the GeV/SiV fluorescence intensity ratio from a nanodiamond, measured by two avalanche photodiodes (APDs). Each data point is calculated by collecting the intensities for 10 s, with 50-ms integration time. The size of the error bars is smaller than that of the markers. The red solid line is an exponential fit. **b**, Long-term time trace of the GeV/SiV fluorescence ratio at room temperature over 25 minutes, showcasing the stability of the measurement. The ratio trace is recorded at 50-ms integration time. **c**, Temperature uncertainty ($\sigma_T$) of the thermometer as a function of integration time. The temperature

resolution is $\eta_T = \sigma_T \sqrt{t_m}$, where $t_m$ is the integration time, and can be extracted from the shot noise fit (red solid line). Each data point is calculated from a total acquisition of 10 s for different integration times. All the measurements were performed using a 400-µW, 637-nm CW excitation laser. **d**, Relative sensitivity plotted versus temperature resolution for several different systems (the resolution on the x-axis is multiplied by the excitation power density): (i) this technique, (ii) intensity of the ZPL of the NV centers in diamond, (iii) wavelength shift of SiV centers in diamond, (iv) linewidth of GeV centers in diamond, (v) fluorescence peaks ratio of NaLuF$_4$:Yb, Er upconversion nanoparticles. Note that techniques vi–x (see main text) are not shown as the power density information is not available (see also Table 1).

**Table 1.** Benchmarking of various nanoscale thermometers on the basis of relative sensitivity, temperature resolution and temperature resolution relative to power density. (i) Our technique, (ii) intensity of the ZPL of diamond NV centers, (iii) wavelength shift of diamond SiV centers, (iv) linewidth of diamond GeV centers, (v) fluorescence peaks ratio of NaLuF$_4$:Yb, Er upconversion nanoparticles, (vi) fluorescence peaks ratio of (Gd,Yb,Er)$_2$O$_3$ upconversion nanoparticles, (vii) wavelength shift of the emission from CdSe quantum dots, (viii) changes in fluorescence lifetime of CdTe quantum dots, (ix) wavelength shift of the emission from triarylboron organic dyes, (x) fluorescence intensity of rhodamine-B organic dyes. (For techniques vi–x, the power density data is not available).

| Group | Thermometers | Rel. Sensitivity (K$^{-1}$) | Temperature resolution (K·Hz$^{-1/2}$) | Temperature resolution relative to power (K·Hz$^{-1/2}$·W·cm$^{-2}$) |
|---|---|---|---|---|
| NDs | GeV/SiV (AS/S ratio) | 0.018 | 0.39 | $5.8 \times 10^4$ |
| | NV (ZPL intensity)[24] | 0.01 | 0.30 | $3.4 \times 10^6$ |
| | SiV (wavelength shift)[30] | $1.61 \times 10^{-5}$ | 0.36 | $1.8 \times 10^6$ |
| | GeV (linewidth)[31] | 0.0064 | 0.30 | $2.5 \times 10^6$ |
| UCNPs | NaLuF$_4$:Yb, Er (PL ratio)[37] | 0.009 | 0.50 | 6.1 × 10$^6$ |
| | (Gd,Yb,Er)$_2$O$_3$ (PL ratio)[38] | 0.015 | 1.00 | — |
| QDs | CdSe (wavelength shift)[19] | 1.61 × 10$^{-4}$ | 1.00 | — |
| | CdTe (PL lifetime)[39] | 0.008 | 0.50 | — |
| ODs | Triarylboron (wavelength shift)[40] | 0.005 | 1.00 | — |
| | Rhodamine-B (PL Intensity)[41] | 0.02 | 0.80 | — |

We now characterize the performance of our ratiometric thermometer. We first established a temperature calibration curve by exciting a representative diamond particle with a 637-nm laser at 400 µW while increasing the temperature from 295.5 to 388 K in steps of 15 K. As the temperature increases, the GeV fluorescence count rate increases, whereas that of the SiV reduces (**Supplementary Figure S2**). **Figure 2a** shows a plot of GeV/SiV fluorescence ratio as a function of temperature (red circles). The data is well fitted with an exponential function $f(T) = y_0 + a * e^{\left(\frac{-b}{T-T_0}\right)}$, where $f(T)$ is the fitting function, $y_0$ is the vertical offset, $T_0$ is the horizontal offset, and $a$ and $b$ are fitting coefficients.[32] From the fit, we obtained a relative sensitivity $S_r = \frac{1}{O}\frac{dO}{dT}$, where $O$ is the measured observable, of 0.018 K$^{-1}$ at 300 K and 0.011 K$^{-1}$ at 388 K. To examine the consistency of the fluorescence ratio, we recorded such a ratio as a function of temperature over the course of 25 minutes as shown in **Figure 2b**. A small ratio fluctuation of 4.5×10$^{-3}$ was extracted from the plot, suggesting the small variance of the measured observable. Furthermore, we tested the repeatability of the sensor by subjecting it to two heating-cooling cycles and monitoring the fluorescence ratio as shown in **Supplementary Figure S3**. The negligible differences between the calibration curves indicate that the thermometer is, in fact, very robust. Notably, the fluorescence ratio varies only slightly as the excitation power is changed (3.7×10$^{-4}$ µW$^{-1}$) as illustrated in **Supplementary Figure S4**.

By varying the integration times of the avalanche photodiodes (APDs) and calculating the standard deviation of the observable, we derived the ratio uncertainty versus integration time. **Figure 2c** shows such relation with a shot noise fit function $\frac{1}{\sqrt{t_m}}$, where $t_m$ is the integration time. The fit was in good agreement with our experimental data, revealing a temperature resolution $\eta_T$=0.39 K·Hz$^{-1/2}$ (at an excitation power of 400 μW). This value is superior to our previous study—where we employed two consecutive (Anti-Stokes and Stokes) excitation cycles to obtain the fluorescence ratio from a single type of color center (GeV) as our temperature observable.[32] In contrast, here we employ a single excitation laser (at 637 nm) to simultaneously excite both the Anti-Stokes (GeV) and Stokes (SiV) emissions. The emissions are acquired in parallel using low-noise, fast APDs which make our current technique superior by any metric, whilst also being simpler.

To benchmark the performance of our nanothermometer against other nanoscale all-optical methods in the literature, we considered the two typical attributes of a thermometer, temperature resolution and relative sensitivity. **Table 1** and **Figure 2d** show a comparison of several representative different systems. Note that **Table 1** has more entries than **Figure 2d**, this is because in **Figure 2d** we plotted on the x-axis the resolution times the power density and the latter is not given in many studies. It is however a crucial parameter as resolution is a relative quantity that increases with signal to noise ratio which is a function of power density as well as integration time. In **Figure 2d**, our method is marked as (i) and the other systems are listed from (ii) to (v). These include techniques based on: (ii) intensity of the ZPL of the NV centers in diamond,[24] (iii) wavelength shift of SiV centers in diamond,[30] (iv) linewidth of GeV centers in diamond,[31] (v) fluorescence peaks ratio of NaLuF$_4$:Yb, Er upconversion nanoparticles.[37] Table 1 additionally includes techniques based on: (vi) fluorescence peaks ratio of (Gd,Yb,Er)$_2$O$_3$ upconversion nanoparticles,[38] (vii) wavelength shift of the emission from CdSe quantum dots,[19] (viii) changes in fluorescence lifetime of CdTe quantum dots,[39] (ix) wavelength shift of the emission from triarylboron organic dyes[40] and (x) fluorescence intensity of rhodamine-B organic dyes.[41] Compared to these systems, our nanothermometer fairs significantly better, both for temperature resolution relative to power density and for relative sensitivity. With reference to **Figure 2d**, the temperature resolution relative to power density of our technique is 5.8×10$^4$ K·Hz$^{-1/2}$·W·cm$^{-2}$, over 30-fold higher than that (1.8×10$^6$ K·Hz$^{-1/2}$·W·cm$^{-2}$) of the next best method (iii). The relative sensitivity of our current method is 0.018 K$^{-1}$, which is also better by a factor ~2 than that (0.01 K$^{-1}$) of the next best method (ii).

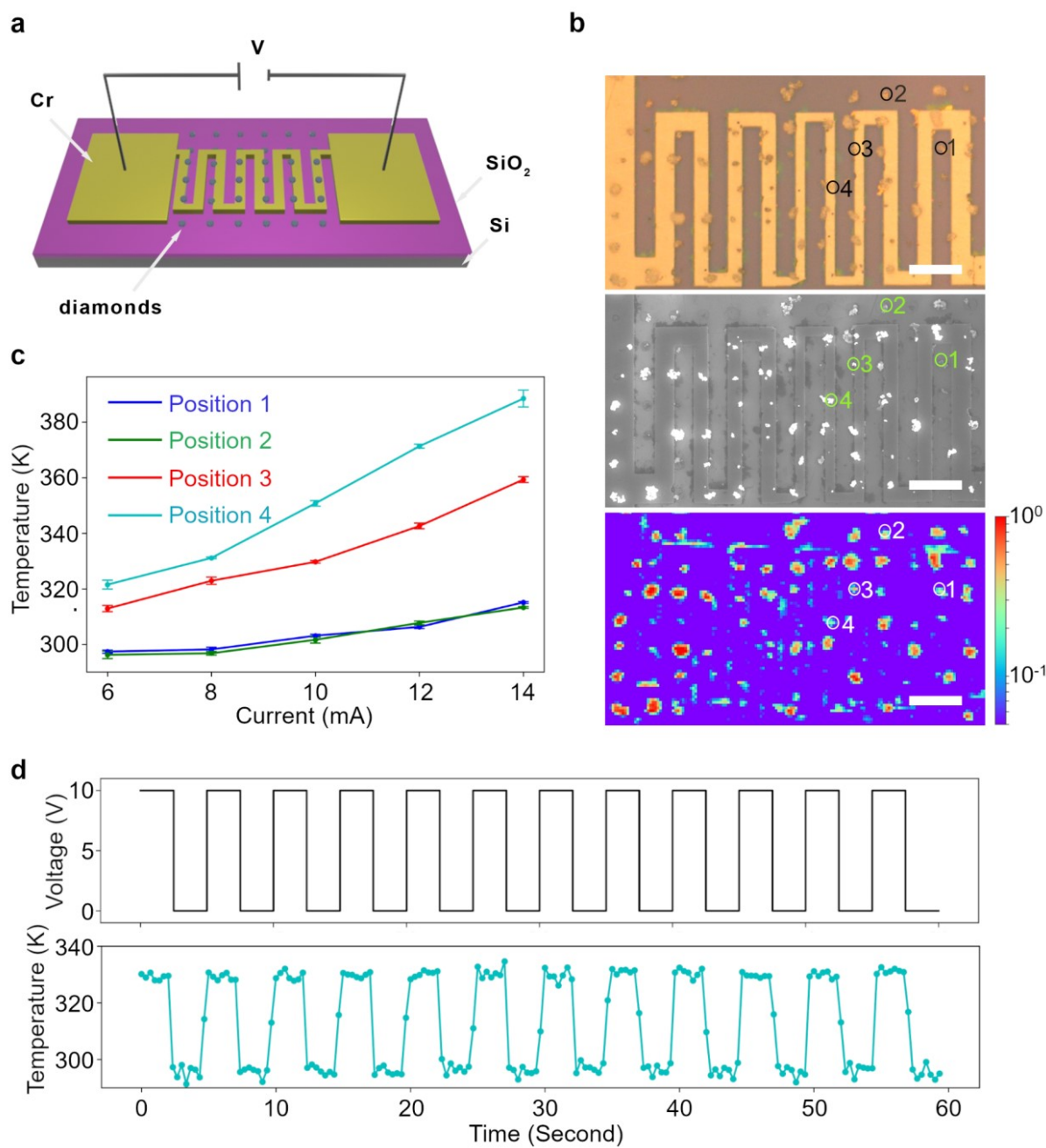

**Fig. 3| Real-time temperature monitoring at localized points of a microcircuit. a**, Schematic illustration of the microcircuit and the array of diamonds nanothermometers. **b**, Optical image, SEM image and corresponding confocal photoluminescence (PL) map of the microcircuit with a 7×13 array of nanodiamonds. The microcircuit was fabricated by electron beam lithography (EBL). The nanodiamond array was produced via capillary assembly of nanodiamonds into an array of 2-μm CSAR (resist) holes patterned by EBL (see main text). A positioning accuracy of ~70% was achieved after removing the polymer resist.[42] The confocal PL map was collected using a 25-μW, 532-nm green CW excitation laser. The inset

scale bar is 10 μm and the color bar of the map is the normalized intensity on a logarithmic scale. **c**, Estimated temperature as a function of input current flown into the microcircuit from four representative temperature sensors whose positions are marked with circles in (b). The temperature readout of each point was calculated from averaging five, 2-s measurements. Each of the 2-s acquisitions was collected at 50-ms APD integration time, under the excitation of 400 μW of a 637-nm CW laser. **d**, Top panel: modulated square-wave voltage (0.2 Hz, with a 50 % duty cycle) as a function of time. Bottom panel: corresponding real-time temperature readouts as a function of time acquired from the sensor in position 4. Each temperature readout was directly extracted from the corresponding fluorescence ratio, with an integration time of 200 ms.

Finally, we demonstrate the temperature sensing capabilities of our thermometer on actual devices, starting with monitoring real-time temperature of a microcircuit. **Figure 3a** shows a schematic of our serpentine-shaped microcircuit with sections of decreasing width from the extremities to the middle: respectively 6 μm, 4 μm and 2 μm. We patterned the microcircuit using electron-beam lithography (EBL) and deposited a thin layer of chromium (200 nm) to form a complete circuit. At micron-sized widths with a serpentine shape, such a device restricts the electron flow, resulting in the formation of local hotspots—an ideal testbed for demonstrating the performance of our nanothermometer.[43] To enable the deposition of nanodiamonds in a patterned array, we employed a capillary-assembly technique (c.f. Methods).[42] Briefly, we patterned an array of 2-μm holes using the CSAR resist; liquid droplets of a nanodiamonds solution were then slowly dragged across the surface of the substrate, forcing the nanodiamonds into the holes with ~70% yield (**Supplementary Figure S5**). **Figure 3b** shows wide-field microscopy, SEM, and confocal microscopy images of the area of interest, where an array of nanodiamonds are clearly visible on the circuit. The array of nanodiamonds allows for pixel-by-pixel thermal mapping of the microcircuit, since all the nanodiamonds are individually calibrated prior to the temperature mapping. To demonstrate the working principle of our approach, we chose four representative nanodiamonds at four different locations (marked with circles). We first tested the circuit by collecting its I-V curve (**Supplementary Figure S6**) which showed the expected linear relationship between input current and voltage (i.e. its Ohmic behavior). Next, we collected individual temperature calibration curves for each selected nanodiamond. The calibration curves were loaded into a script so that, during an actual measurement, the GeV(Anti-Stokes)/SiV(Stokes) intensity ratios could be translated—in real-time—to corresponding temperature values. **Figure 3c** features a plot of predicted temperatures as a function of the input current (6–14 mA) in the microcircuit, based on the four target nanodiamonds. As the nanodiamond in position 4 lied directly above the 2-μm section of the circuit, it exhibited the largest change (68 K) in temperature, increasing from 321 to 389 K. The nanodiamond in position 3 was situated at the edge of the circuit, and thus experienced a smaller temperature increase (46 K) compared to that of the nanodiamond in position 4. The other two nanodiamonds, at position 1 and 2, were ~2–3 μm away from the circuit, and hence showed the smallest temperature difference (17 K). Next, to demonstrate real-time monitoring of temperature at a particular point of the microcircuit, we chose the nanodiamond at position 4 as a case study. By using a signal generator, we modulated the input voltage between 0 and 10 V, in a square-wave pattern at 0.2 Hz and 50% duty cycle, as shown in **Figure 3d**, upper panel. **Figure 3d**, lower panel, shows the corresponding real-time temperature monitoring of the nanodiamond at position 4. It is clear that the temperature switches between 295 and 330 K, as the voltage is accordingly modulated between 0 and 10 V. The fast response indicates that our thermometer is well-suited for nanoscale real-time temperature monitoring.

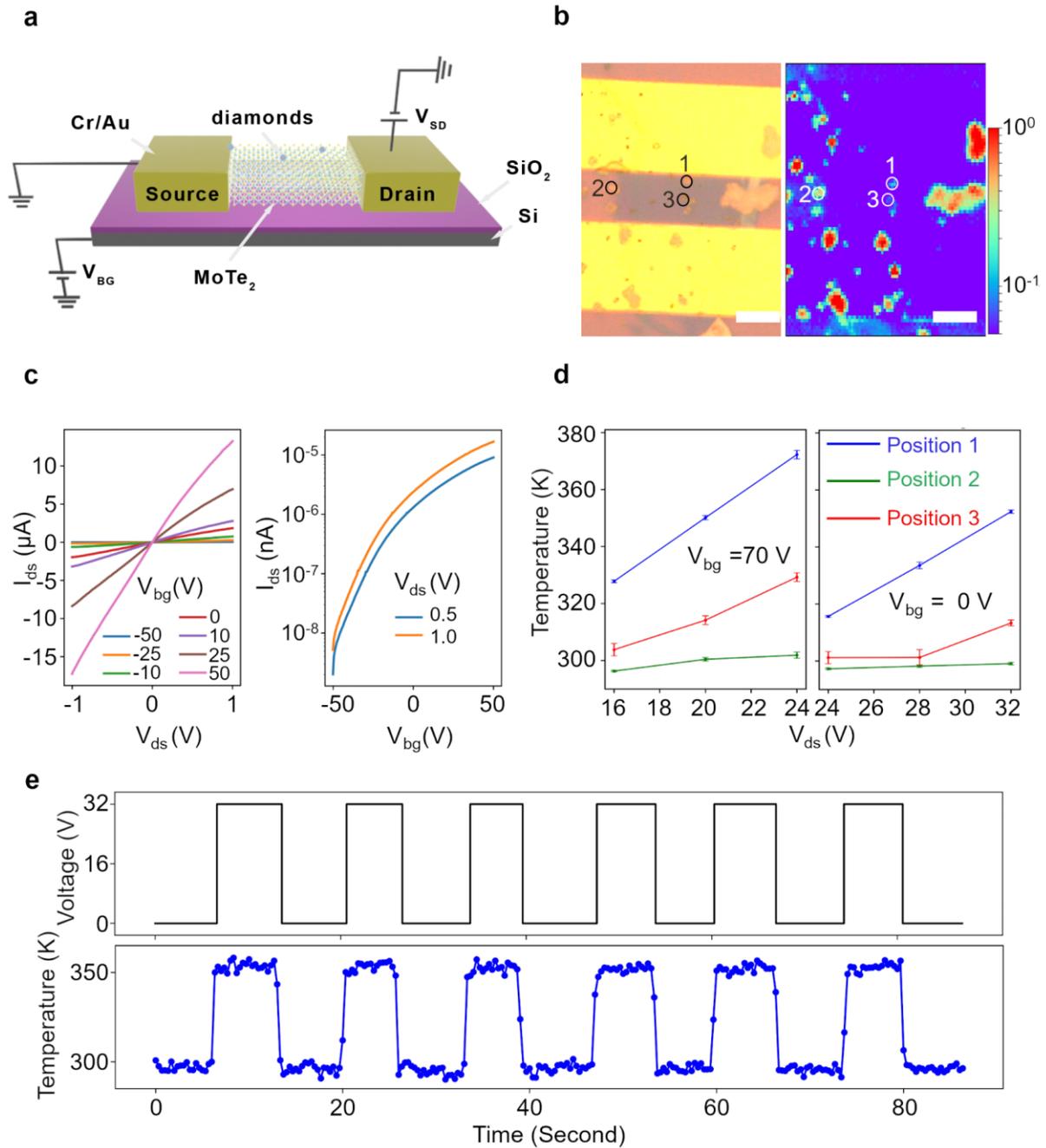

**Fig. 4| Real-time temperature monitoring in a MoTe₂ field-effect transistors (FET). a**, Schematic illustration of the MoTe₂ FET and of the diamond nanothermometers. **b**, Optical image and corresponding confocal PL map of the device consisting of a pair of Cr/Au (5 nm/80 nm) electrodes deposited on a multilayer MoTe₂ flake. Three representative nanodiamonds at three different locations (black and white circles) chosen for temperature monitoring. The inset scale bar is 5 μm and the color bar of the map is the normalized intensity on a logarithmic scale. **c**, Drain-source I-V curves of the MoTe₂ FET under different backgate voltages (left panel), and transfer characteristics of the MoTe₂ FET under the drain-

source voltage of 0.5 V and 1 V, on a logarithmic y-axis scale (right panel). **d**, Temperature readouts plotted versus input drain-source current for the nanodiamonds at three selected locations (shown in (b)), in the MoTe$_2$ FET under a backgate voltage of 70 V (left panel) and 0 V (right panel). Each temperature readout was calculated from five sets of 2-s acquisitions. The integration time was 50 ms under the excitation of 500 $\mu$W of a 637-nm CW laser. **e**, Upper panel: modulated square-wave voltage as a function of time. Lower panel: the corresponding real-time temperature readout as a function of time acquired from the nanodiamond in position 1. The applied drain-source voltage was 32 V under the backgate voltage of 0 V. Each temperature readout was calculated from the fluorescence ratio, with an integration time of 200 ms. We also show that one could trade off temperature resolution for integration time and vice versa, as shown in **Supplementary Figure S7**.

To further test our thermosensor, we performed real-time temperature monitoring on a working field-effect transistor (FET). Specifically, we chose molybdenum ditelluride (MoTe$_2$) as it is a good example of a two-dimensional semiconductor. Such field-effect transistors made from transition metal dichalcogenides (TMDs) have been of intense research interest recently, owing to their potential in the manufacturing of ultra-thin wearables[44,45] and high-density vertical electronics.[46-48] Heat distribution within these extremely thin devices has, however, remained poorly understood.[49] To fabricate the device, we first exfoliated MoTe$_2$ onto a 300-nm thick SiO$_2$-on-silicon substrate and selected the thin flakes (~5 nm). The electrode patterns in contact with the MoTe$_2$ flake were made using EBL, and thin layers of chromium (5 nm) and gold (80 nm) were deposited to make the drain and source electrodes, as shown in **Supplementary Figure S8a**. We used Raman spectroscopy to confirm the selected MoTe$_2$ was of high-quality (**Supplementary Figure S8b**) and employed atomic force microscopy (AFM) to determine the thickness of the flake to be ~5 nm. We then drop-casted the nanodiamonds on the transistor and used them to measure local temperature in real-time. **Figure 4a** shows a simplified schematic of our FET device, where the drain and source electrodes are in good contact with the MoTe$_2$ flake, and the bottom silicon acts as the backgate to exert an out-of-plane electric field on the flake—controlling the current flow across the device. As shown in the optical image and confocal map in **Figure 4b**, there were several nanodiamonds in the region of interest of the device. We selected three representative nanodiamonds at position 1, 2, and 3 to monitor the local temperature of the device. To ensure that the device behaved as a field effect transistor, we first obtained the drain-source $I_{ds}$-$V_{ds}$ curves at different backgate voltages, $V_{bg}$, as shown in **Figure 4c**, left panel. The larger $V_{bg}$ resulted in a steeper $I_{ds}$-$V_{ds}$ slopes—in good agreement with previous works.[50,51] We also tested the transfer characteristics of the transistor at 0.5 and 1 V of the drain-source voltage $V_{ds}$ (**Figure 4c**, right panel) and found that such behaviors were in accordance with earlier reports on this type of transistor.[50,51]

We now turn to the temperature monitoring of the transistor using the nanodiamonds. **Figure 4d** shows two plots of local temperature as a function of $V_{ds}$, with two different $V_{bg}$ at 70 V and 0 V, for the three nanodiamonds. We first examined the plot with $V_{bg}$ = 70 V (left panel). The diamond at position 1, situated at the boundary of the electrode and the MoTe$_2$, exhibited the largest increase in temperature, 44 K, from 328 to 372 K, when $V_{ds}$ changed from 16 to 24 V. Such a large temperature increase can be attributed to the Schottky barrier at the metal-semiconductor junction, resulting in massive heat accumulation in this area. The nanodiamond located at position 3 experienced significantly less heating, with only a 25 K increase in temperature under the same conditions. The nanodiamond at position 2 experienced a relatively modest increase in temperature, 6 K, consistent with its location at the boundary between the MoTe$_2$ flake and the substrate. With $V_{bg}$ = 0 V (right panel), a similar trend was observed for the three nanodiamonds, allowing us to extract information about the local heat distribution pattern on the transistor, while in operation. Finally, to monitor the real-time temperature of the device at position 1, we generated a modulated signal of $V_{ds}$ between 0 and 32 V with the fixed $V_{bg}$ at 0 V (**Figure 4e**, top panel), and recorded the temperature of the nanodiamond as a function of time. As shown in **Figure 4e**, bottom panel, the changes in temperature were in accordance with that of the voltage modulation, showing a 56 K difference in temperature between 0 and 32 V of $V_{ds}$.

To conclude, we demonstrated a real-time nanothermometry approach that utilizes the fluorescence intensity ratio between GeV (Anti-Stokes regime) and SiV (Stokes regime) in nanodiamonds engineered to be co-doped with both centers. We show that the nanothermometer has excellent sensitivity (0.018 K$^{-1}$ at 300 K) and temperature resolution relative to power density ($5.8 \times 10^4$ K·Hz$^{-1/2}$·W·cm$^{-2}$). These are a factor ~2× and ~30× better, respectively, than those of the next best all-optical nanothermometers. We demonstrated the suitability of our nanothermometer for fast, real-time monitoring of local temperature in a microcircuit and a MoTe$_2$ field-effect transistor. When combined with the deterministic deposition of arrays of nanodiamonds, pixel-by-pixel thermal mapping can be achieved in these devices. Our work paves the way to fast temperature monitoring and mapping of micro-and nano-scale devices such as nanoelectronic circuits, microfluidic channels and nanophotonic devices, as well as of complex biological environments like tissues and cells.

## Methods

### Sample preparation

Ge- and Si- co-doped nano-/micro-diamonds were produced at high pressures and temperatures in C-H-Si(0.19 at%)-Ge(0.2 at%) growth system. Micro-diamonds were synthesized at a pressure of 8 GPa and a temperature of 1800–2000 °C, while nanodiamonds at higher pressure of 9 GPa and reduced temperature of 1500–1600 °C for about 60 s. For the synthesis experiments, powder mixture of Adamantane $C_{10}H_{16}$ (300 mg, > 99%, Sigma-Aldrich), Tetraphenylgermane $C_{24}H_{20}Ge$ (15mg, 96%, Aldrich) and Tetraphenylsilane $C_{24}H_{20}Si$ (18mg, 96%, Aldrich) were mixed in a mortar and pestle, both made of jasper, for about 5 min, pressed into a pellet (65 mg) and placed inside a titanium capsule (6 mm in diameter, 4 mm in height, with 0.2-mm wall thickness). A toroid-type, high-pressure chamber was used to generate pressure and temperature in the reaction cell.[34] After the treatment, samples were quenched under pressure to room temperature.

The nanodiamonds were then dispersed in isopropanol (IPA) at the concentration of 0.1% (w/w). The size of the nanodiamonds were ~300-500 nm. Five microliters of the solution were drop-cast on a clean 0.5 x 0.5 $cm^2$ silicon substrate and left to dry on a hotplate at 60 °C to completely remove the residual solvent. The silicon chip was then ready to be used for optical or structural characterization.

### Device fabrication

**Preparation of Microcircuits**. A volume of ~0.1 mL of polymethyl methacrylate resist solution (950 PMMA) was spun-cast on the thermally-grown $SiO_2$ (~300 nm) on Si substrate for 1 min at 3000 rpm to get a resist coating with the thickness of ~200 nm. The resist coating was then patterned using a scanning electron microscopy (SEM) (Zeiss Supra 55VP) coupled to the Raith electron beam lithography (EBL) system. The resist pattern was subsequently formed by immersing in the resist developer (a methyl isobutyl ketone (MIBK)/isopropyl alcohol (IPA) (1:3) solution) for 30 seconds and the resist stopper (IPA) for 1 min. After cleaning the resist residual of the pattern area via $O_2$ plasma for 10 s under 50 sccm $O_2$ and 100 W power, 200 nm of chromium was deposited onto the patterned resist coating using a lab-built plasma-assisted sputter deposition chamber. The microcircuit was obtained by immersing into the resist remover (99%, acetone) to eliminate the remaining resist and excessive metal.

**Preparation of MoTe$_2$ transistor.** MoTe$_2$ flakes were mechanically exfoliated from bulk MoTe$_2$ crystal onto a silicon substrate with a 285-nm thermally grown silicon dioxide layer using the scotch tape method.[52] A few-layer MoTe$_2$ flake (<10 nm) was selected based on its optical contrast and its thickness was measured using an atomic force microscopy (AFM, Park XE7). A 2-hour, 200 °C annealing process under 50 sccm Ar was adopted to remove the excessive contamination of the flake source and increase the interaction force between the substrate and the flake. Furthermore, two electrode contacts (source and drain) aligned with the targeted flake were formed via electron beam lithography. First, a 300-nm positive e-beam resist (CSAR 13.5%) was coated to the substrate at a spin rate of 4000 rpm for 1 minute, followed by 170 °C baking on a hotplate to remove the residual solvent. For aligning the pattern and targeted MoTe$_2$ flake, focused electron beams (20 keV) at very low beam current (37 pA) were used to avoid exposure of the unpatterned area. Then the pattern structures were irradiated at a dosage of 150 μC/cm$^2$. The source-drain channel width is 6 μm. After a development process, 5 nm of chromium and 80 nm of gold were deposited by thermal evaporation deposition. The remaining metals were removed in the lift-off process. Ten microliters of nanodiamonds solution were drop-cast onto the substrate to deposit a few nanodiamonds on the transistor and let it dry naturally in air.

**Preparation of an array of nanodiamonds on microcircuit.** To deposit diamond particles on the microcircuit, a second EBL patterning step was performed to create an array of open apertures. Ten microliters of nanodiamonds aqueous solution (0.5% w/w) containing 0.01% CTAB (Hexadecyltrimethylammonium bromide) surfactant were drop-casted on the substrate and a thin coverslip was slowly glided at an angle across the substrate to assist the insertion of the nanodiamonds into the apertures. The substrate was then dried naturally in air. The details of the process can be found elsewhere.[42]

**Optical characterization**

Optical characterization of the quantum emitters was conducted using a lab-built confocal microscope. The sample was glued using silver paste onto the high-precision temperature-controlled stage (Microptik, MHCS-622) with a temperature resolution of 0.1 K. A 637-nm continuous wave (CW) laser beam was focused onto the diamond particles through a long-working-distance objective with a numerical aperture of 0.7 (Mitutoyo, Plan Apochromat, 100×). A fast-steering mirror (Newport, FSM-300-01) was used to control the laser spot position and scan across the sample to generate a confocal map. The excitation laser and the photoluminescence (PL) signals were separated by a plate beam splitter (70T:30R) (Thorlabs). The collection arm was further split into two paths, one for the GeV emission (anti-Stokes PL) and the other for the SiV emission (Stokes PL). To completely suppress the 637-nm excitation laser, we inserted a tunable notch filter (Semrock, NF03-642E-25) in the main collection arm, and a tunable shortpass filter (Semrock, TSP01-628-25x36) for the GeV collection path and a tunable longpass filter (Semrock, FF01-715/LP-25) for the SiV counterpart. Each collection arm was fiber-coupled into a graded-index multimode fiber (Thorlabs, GIF625) connected to a single-photon avalanche photodiode (SPAPD) (Excelitas Technologies, SPCM-AQRH-TR). Software was developed (LabView) to control all the hardware, analyze the photon rates and measure the real-time fluorescence ratio/temperature based on the individual signals collected at each APD. Simultaneously, the photoluminescence from individual nanodiamonds was also spectrally analyzed with a spectrometer (Princeton, Acton Series SP2300 series) equipped with a charge-coupled device (CCD) camera (PIXIS series). In most of the optical measurements, the 637-nm laser was pumped at 400 μW of power, measured at the back aperture of the objective. To acquire the full spectra of both GeV and SiV emission, we employed a 532-nm laser with 50 μW of power.

## Author contribution

Y. C. and T. T. T conceived the idea of the project. Y.C. and T. T. T built the optical system and its software. E. E. fabricated the co-doped nanodiamonds. Y. C. made the microcircuit, performed the optical characterization and the thermometric measurements. C. L., T. Y. and Y. C. fabricated the transistor. Y.C. and T. T. T analyzed the data. T. T. T supervised the project. All authors discussed the results and commented on the manuscript.

## Fundings

We acknowledge the Australian Research Council (CE200100010, DP190101058) and the Asian Office of Aerospace Research and Development (FA2386-20-1-4014) for the financial support. E.A. Ekimov thanks the support from the Russian Science Foundation, Grant No. 19-12-00407. C Bradac thanks NSERC (RGPIN-2021-03059) and CFI JELF (#41173) for financial support.

## Acknowledgement

We thank Dr. Mehran Kianinia for fruitful discussion and technical assistance with building the optical system and its software and Dr. Zai-Quan Xu for technical advice on the fabrication of the transistor device and help with AFM measurement. The authors thank the ANFF (UTS node) for use of the diamond reactor facilities.

## Notes

The authors declare no competing financial interest.